\documentclass[twocolumn]{article}
\usepackage[pdftex]{graphicx}
\makeatletter
\renewcommand{\@cite}[1]{\textsuperscript{#1)}}
\def\@biblabel#1{#1)}

\makeatother
\usepackage{geometry}
\usepackage{caption}
\begin{document}

\onecolumn

\title{\bf{Electrical and Thermal Properties of \\SnTe/Sb$_{2}$Te$_{3}$ Superlattice Phase Change Materials}}
\author{Toshimichi Shintani}
\maketitle
\begin{center}
{\it Collaborative Research Team, Green Nanoelectronics Center\\National Institute of Advanced Industrial Science and Technology}

({\it Current affiliation: CT Applied Science Office\footnote{email address: ctappsci(at)gmail.com}})

\end{center}

\begin{abstract}
The fundamental electrical and thermal properties of the devices consisting of Sn$_{x}$Te$_{1-x}$/Sb$_{2}$Te$_{3}$ superlattice (SnTeSL) materials have been investigated and compared with those of the conventional Ge$_{2}$Sb$_{2}$Te$_{5}$ (GST225) and GeTe/Sb$_{2}$Te$_{3}$ superlattice (GeTeSL) in terms of their resistance switching characteristics. A significant reduction in the switching power of SnTeSL is demonstrated by conducting a proper initialization procedure, varying {\it x} in Sn$_{x}$Te$_{1-x}$/Sb$_{2}$Te$_{3}$, and applying short electric pulses. It is found that the observed drastic power reduction occurs due to the exponential decrease in the electric current under the pulse incidence. On the other hand, the thermal properties of the studied SL materials are very similar to those of the conventional phase change materials. The obtained transmission electron micrographs and results of multilevel-cell recording in GeTeSL and SnTeSL via multi-pulse incidence are totally different from the properties of  the phase change materials along the GeTe-Sb$_{2}$Te$_{3}$ pseudo-binary tie line. Two different models (the partial switching and electric field-induced one) are proposed to elucidate the mechanism of the resistance switching in the SL materials.
\end{abstract}

\twocolumn

\section{Introduction}
\label{Introduction}

Phase-change memory (PCM)\cite{ovshinsky}, which has been practically implemented in rewritable optical disks, is a promising candidate for the next-generation solid-state memory technologies\cite{burr}. It possesses many advantages such as excellent high-speed switching characteristics and good scalability, which can be potentially utilized for the fabrication of high-speed large-capacity memory chips. However, PCM also has drawbacks, including high power consumption during the resistance switching from the low to the high state (reset operation), which is required for melting the PC component. If the melting point of the utilized material (and, therefore, the crystallization temperature) is too low, its reset power {\it P}$_{rst}$ becomes very small, leading to the deterioration of its retention properties. Among various PC materials, Ge$_{2}$Sb$_{2}$Te$_{5}$ (GST225)\cite{yamada}\cite{wuttig} possesses a relatively low melting point, adequate crystallization characteristics, and good overwritability (endurance properties). However, even this material is unable to fully satisfy the requirements for PCM devices\cite{burr}.
In previous studies, a superlattice-like (SLL) PC material consisting of alternately stacked GeTe and Sb$_{2}$Te$_{3}$ layers has been successfully prepared\cite{chong}\cite{zhao} and utilized for the reduction of {\it P}$_{rst}$ to a magnitude equal to about 1/3 the GST225 level. Surprisingly, this material demonstrated high overwritability despite the expectations that the layered structure would be destroyed after multiple melting cycles. Furthermore, the superlattice (SL) PC material has been proposed\cite{tominaga}\cite{simpson}  (“GeTeSL”). While this SL consists of the alternate stacks of GeTe and Sb$_{2}$Te$_{3}$ as SLL, its {\it P}$_{rst}$ magnitude is equal to 1/10 of the GST225 value, which is about three times lower than that of SLL. The studied SL and SLL materials mainly differ in their layer thicknesses and crystalline structures. Thus, the thickness of the GeTe layer in SL is about 1 nm, while that in SLL is about 3.5 nm. Similarly, the crystalline structure of each GeTe layer in SL is aligned along the (111) direction\cite{soeya1}, whereas the SLL layers are randomly oriented. However, these differences cannot fully elucidate the mechanism of lowering {\it P}$_{rst}$ in SL and SLL. After the proposal of GeTeSL, a number of studies\cite{shintani}$^{-}$\cite{ohyanagi3} have been performed to investigate the properties of GeTeSL. Some studies\cite{kalikka}\cite{zhou} have proposed the mechanism of low power switching in SL materials, which, however, should remain investigated more on the possibility to explain all the experimental results on SL materials including the results to be reported in this paper.

Moreover, a much lower {\it P}$_{rst}$ has been obtained for the SL structure consisting of SnTe and Sb$_{2}$Te$_{3}$ layers ("SnTeSL")\cite{soeya2}. In particular, the {\it P}$_{rst}$ of Sn$_{0.1}$Te$_{0.9}$/Sb$_{2}$Te$_{3}$ SL is smaller than that of GST225 by a factor of more than 10$^{3}$. However, the reasons for such low {\it P}$_{rst}$ and fundamental properties of this material remain unclear.

In this work, the fundamental characteristics of SnTeSL are examined experimentally, and the mechanism of the resistance switching in the SL PC materials is discussed by comparing them with the properties of the GST225 and GeTeSL materials. As a result, a reduction in {\it P}$_{rst}$ by a factor of 10$^{-3}$ is observed for Sn$_{x}$Te$_{1-x}$/Sb$_{2}$Te$_{3}$ SL with $x = 0.1$ as compared to the value obtained for GST225. Experimental studies are performed by initializing the SL materials with short electric pulses; investigating their scaling and annealing properties; taking transmission electron microscopy (TEM) images of the as-deposited, set, and reset states; and obtaining multilevel cells (MLC) by applying multiple pulses.

\section{Experimental}
\label{Experimental}

\begin{figure}[hbt]
 \begin{center}
  \includegraphics[clip,width=2in]{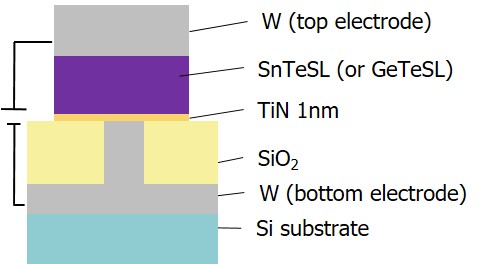}
  \caption{Cross-sectional view of the devices used in this study.}
  \label{fig1}
 \end{center}
\end{figure}

SnTeSL and GeTeSL samples were prepared using the procedure described in the previous reports\cite{soeya1}\cite{shintani}\cite{soeya2}. Their cross-sections are depicted in Fig.\ \ref{fig1}. The utilized substrates consisted of the W electrodes deposited on Si wafers below SiO$_{2}$ layers. The diameter of the electrode in this study was 120 nm if not specified otherwise. The remaining device structure consisted of TiN (1 nm), Sb$_{2}$Te$_{3}$ (10 nm), PC, and W (50 nm) films obtained by sputter deposition. The PC films were GST with thicknesses of 50 nm: [GeTe (1 nm)/Sb$_{2}$Te$_{3}$ (4 nm)]$_{8}$ for GeTeSL and [Sn$_{x}$Te$_{1-x}$ (1 nm)/Sb$_{2}$Te$_{3}$ (4 nm)]$_{8}$ for SnTeSL. To deposit a SL film, GeTe, SnTe, and Sb$_{2}$Te$_{3}$ targets were sputtered alternately. During sputtering, the substrate was heated to 250$^\circ$C in the cases of GST and GeTeSL, and to 200$^\circ$C in the case of SnTeSL. The reason for selecting a lower temperature for SnTeSL was to avoid alloying the SnTe and Sb$_{2}$Te$_{3}$ layers. After sputtering, electronic devices with sizes of about 100 $\mu$m were patterned using photolithography and reactive ion etching techniques.

\begin{figure}[hbt]
 \begin{center}
  \includegraphics[clip,width=3.4in]{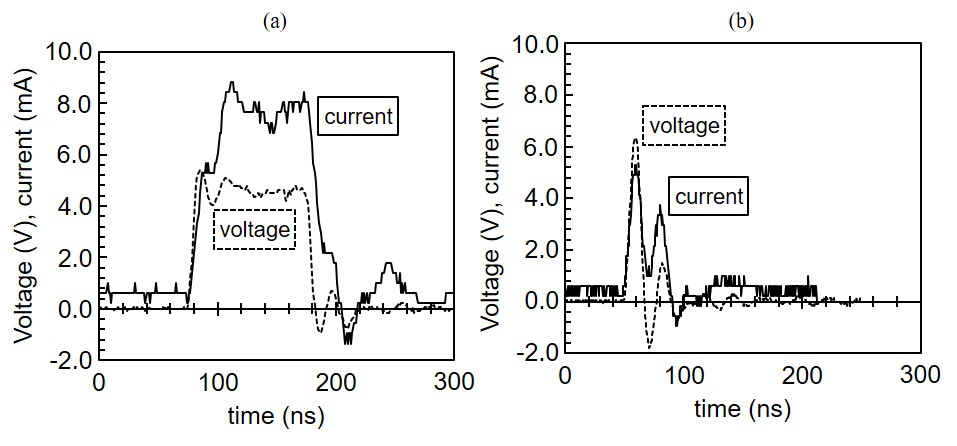}
  \caption{Shapes of the electric current and voltage pulses passing through the tested devices, which were measured using the active FET probe. The programmed pulse lengths were (a) 100 ns and (b) 10 ns.}
  \label{fig2}
 \end{center}
\end{figure}

The electric properties of the fabricated devices were investigated using the tester described elsewhere\cite{shintani}, which applied write and read pulses to the probes connected to the bottom and top electrodes. The read voltage was 0.1 V. The shape of the programmed pulse was rectangular whose typical length was 100 ns if not specified otherwise. The voltage pulses applied to the tested device became dull with time due to the existence of parasitic resistance and capacity, which were measured using an active field-effect transistor (FET) probe. Fig.\ \ref{fig2} shows the time evolution of the voltage and current passing through the device, indicating that the rise and fall times of the generated pulses ranged between 5 and 10 ns. 

\begin{figure}[hbt]
 \begin{center}
  \includegraphics[clip,width=3.4in]{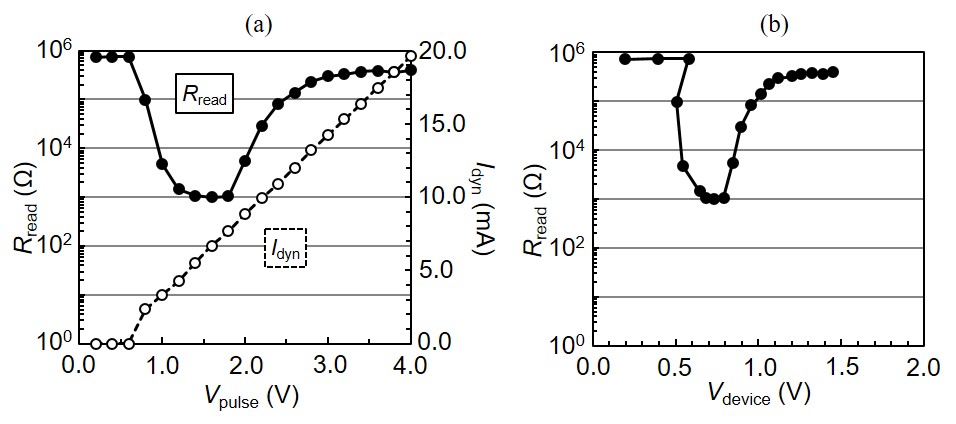}
  \caption{Relationships between the applied voltage and resulting read resistance {\it R}$_{read}$ (corresponding to the dynamic current {\it I}$_{dyn}$) obtained for the GST device. The vertical axes denote the (a) programmed and (b) practically applied voltages.}
  \label{fig3}
 \end{center}
\end{figure}

The electrical properties of the GST device measured using this system are shown in Fig.\ \ref{fig3}, where {\it V}$_{pulse}$ is the pulse voltage set by the computer, and {\it V}$_{device}$ denotes the voltage measured with the active FET probe during the pulse incidence. The differences between the data presented in Fig.\ \ref{fig3}(a) and (b) result from the dependence of the device resistance on the applied voltage. The total resistance in the measurement system can be written as the sum of the device resistance {\it R}$_{device}$ and the parasitic resistances brought by other components {\it R}$_{para}$, {\it i.e.}, {\it R}$_{device}$ + {\it R}$_{para}$. Accordingly, we can write {\it V}$_{device}$ = {\it V}$_{pulse}$ × {\it R}$_{device}$ / ({\it R}$_{device}$ + {\it R}$_{para}$). Since {\it R}$_{para}$ can be considered independent of the pulse voltage and the device resistance, {\it V}$_{device}$ rapidly decreases when {\it R}$_{device}$ decreases rapidly by crystallization of GST225 in the device. 

The thermal properties of the SnTeSL and GeTeSL devices were investigated by placing them on a hot plate and measuring their read resistances while increasing the plate temperature at a heating rate of 1$^\circ$C/s.
TEM observations were performed at Toray Research Center (Japan). The utilized TEM samples were Sn$_{0.5}$Te$_{0.5}$/Sb$_{2}$Te$_{3}$ devices in the initial, initialized low resistance (set) and reset states, and the broken state. The reset state was obtained after 10 rewriting cycles, whereas the broken cell was prepared by applying a high reset voltage of 4 V instead of the typical value equal to around 1 V.

\section{Results}
\label{Results}

\subsection{Fundamental properties of the Sn$_{0.5}$Te$_{0.5}$/Sb$_{2}$Te$_{3}$ SL device}
\label{fundamental}

\begin{figure}[hbt]
 \begin{center}
  \includegraphics[clip,width=3.4in]{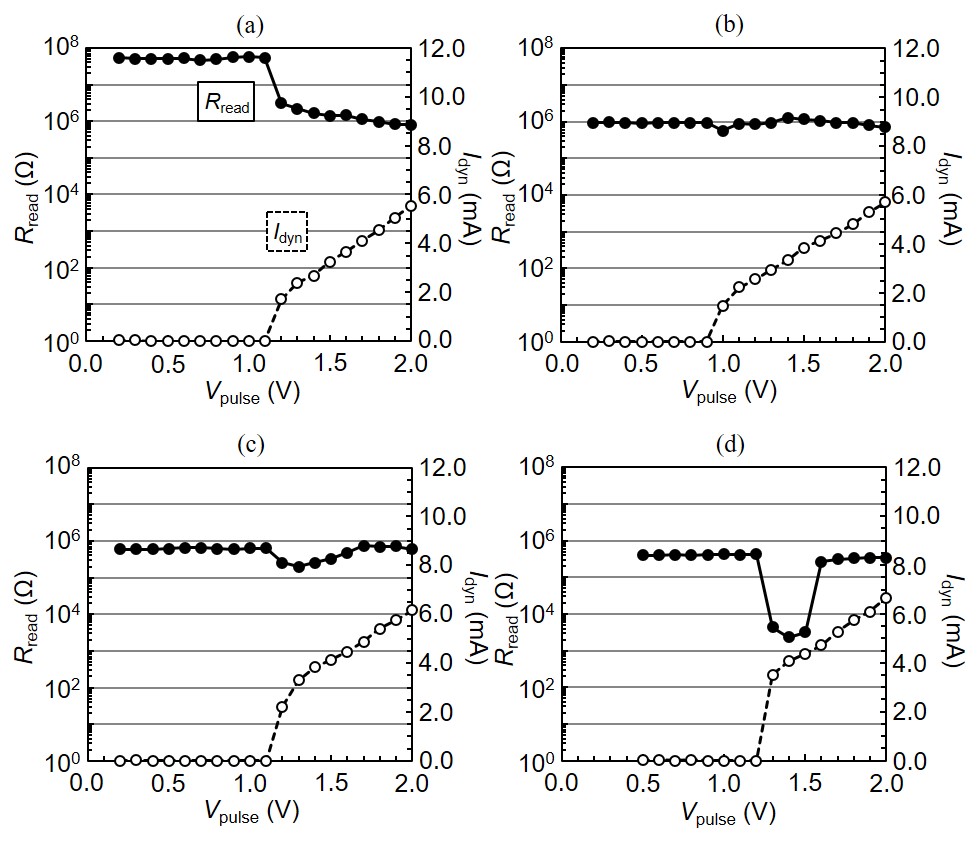}
  \caption{Initialization of the Sn$_{0.5}$Te$_{0.5}$/Sb$_{2}$Te$_{3}$ SL material with high initial resistance. The applications of the (a) first, (b) second, (c) third, and (d) fourth pulses.}
  \label{fig4}
 \end{center}
\end{figure}

Fig.\ \ref{fig4} describes the initialization process of the prepared Sn$_{0.5}$Te$_{0.5}$/Sb$_{2}$Te$_{3}$ SL device. It shows that its initial state corresponds to the high resistance state, although all the films inside the activated region retained their crystalline properties\cite{soeya2}. This high initial resistance was observed in the majority of the cells in this study and also in GeTeSL. This property is similar to the previous reports\cite{simpson}. For the reasons currently remaining unclear, the initial resistances of most devices with low switching power are typically high, which will be discussed later in this work. Furthermore, the obtained results reveal that successful resistance switching requires the application of several initializing pulses. The initialization experiments conducted using multiple cells show that the number of these pulses depends on the cell type and is typically varied between one and three, while the utilized pulse voltage lies in the region of 1.0 - 1.2 V.

\begin{figure}[hbt]
 \begin{center}
  \includegraphics[clip,width=3.4in]{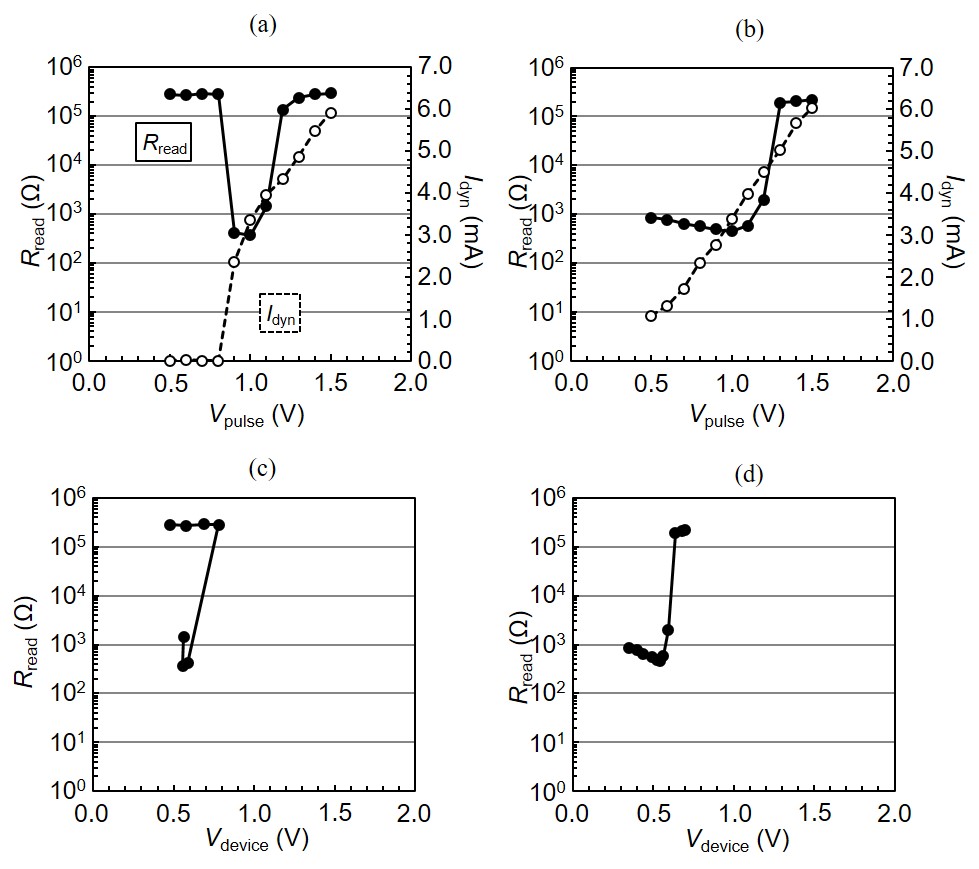}
  \caption{Relationships between the voltage and read resistance {\it R}$_{read}$ (corresponding to the dynamic current {\it I}$_{dyn}$) obtained for the Sn$_{0.5}$Te$_{0.5}$/Sb$_{2}$Te$_{3}$ SL device. The initial states are the (a)(c) high and (b)(d) low resistance states. The vertical axes denote the (a)(b) programmed voltage Vpulse and (c)(d) practically applied voltage {\it V}$_{device}$.}
  \label{fig5}
 \end{center}
\end{figure}

Fig.\ \ref{fig5} shows the results of the pulse sweep measurements performed after the initialization process. The obtained reset power was about 1/15 of that of the GST device. The data presented elsewhere\cite{shintani} show that the {\it P}$_{rst}$ of the Ge$_{0.5}$Te$_{0.5}$/Sb$_{2}$Te$_{3}$ SL device was about 1/10 of value obtained for the GST device, suggesting that the {\it P}$_{rst}$ of SnTeSL was of the same order of magnitude as that of GeTeSL.

\begin{figure}[hbt]
 \begin{center}
  \includegraphics[clip,width=3.4in]{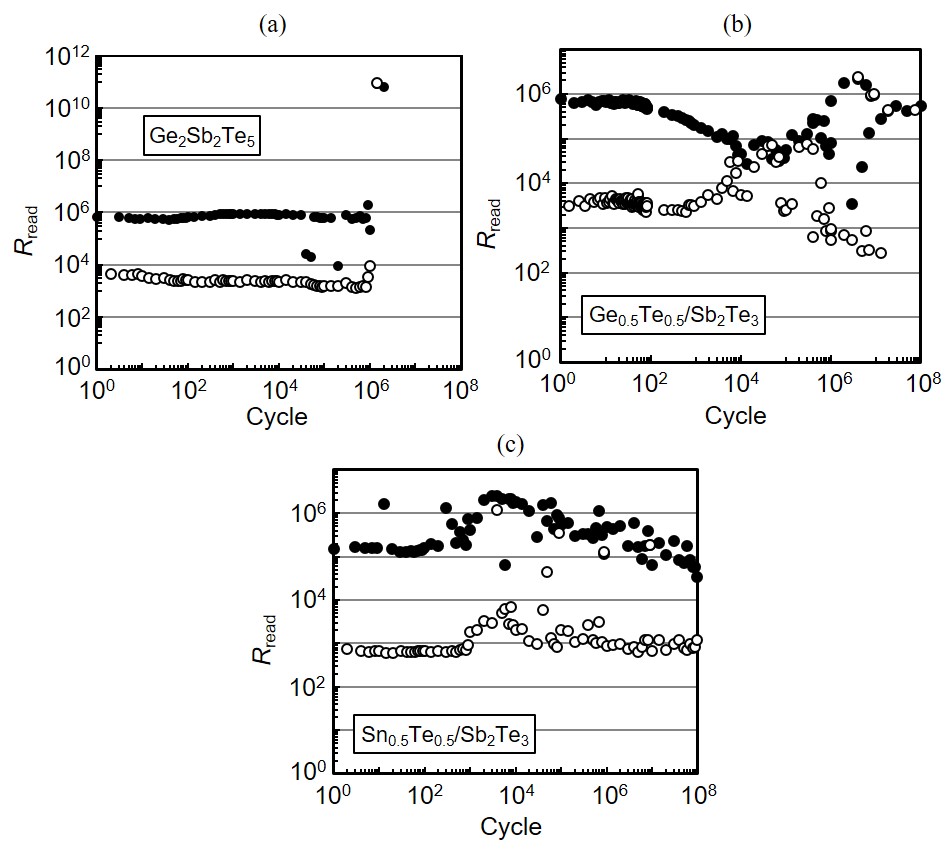}
  \caption{Endurance properties of the (a) GST, (b) Ge$_{0.5}$Te$_{0.5}$/Sb$_{2}$Te$_{3}$ SL and (c) Sn$_{0.5}$Te$_{0.5}$/Sb$_{2}$Te$_{3}$ SL materials.}
  \label{fig6}
 \end{center}
\end{figure}

Fig.\ \ref{fig6} describes the endurance properties of the SnTeSL device with the endurance properties of GST and GeTeSL devices for comparison. This figure indicates that it is capable of sustaining at least 10$^{8}$ switching cycles in SL devices, which confirm the results almost equivalent to the previous one reported by Simpson {\it et al.}\cite{simpson}. In contrast, the maximum number of cycles obtained for the GST device was 10$^{6}$. The better endurance performance of SL devices may result from its lower switching power, which generates a lower amount of stress inside the cell. However, the read resistances of SL devices are much less stable than those of GST. Both the set and reset resistances increase after 10$^{3}$ – 10$^{4}$ or 10$^{5}$ cycles and recover their original values after overwriting. While this feature is undesirable for practical use, the relative instability of the read resistance can be corrected by controlling the set and reset voltages during rewriting (data not shown). Thus, this instability can be solved by verification of the formed read resistance which has been already adopted in the current semiconductor memories such as flash memories. This instability, however, should be solved if solvable. To this end, the cause of this instability should be clarified, which will be discussed in Sec. \ref{discussions}.

\begin{figure}[hbt]
 \begin{center}
  \includegraphics[clip,width=3.4in]{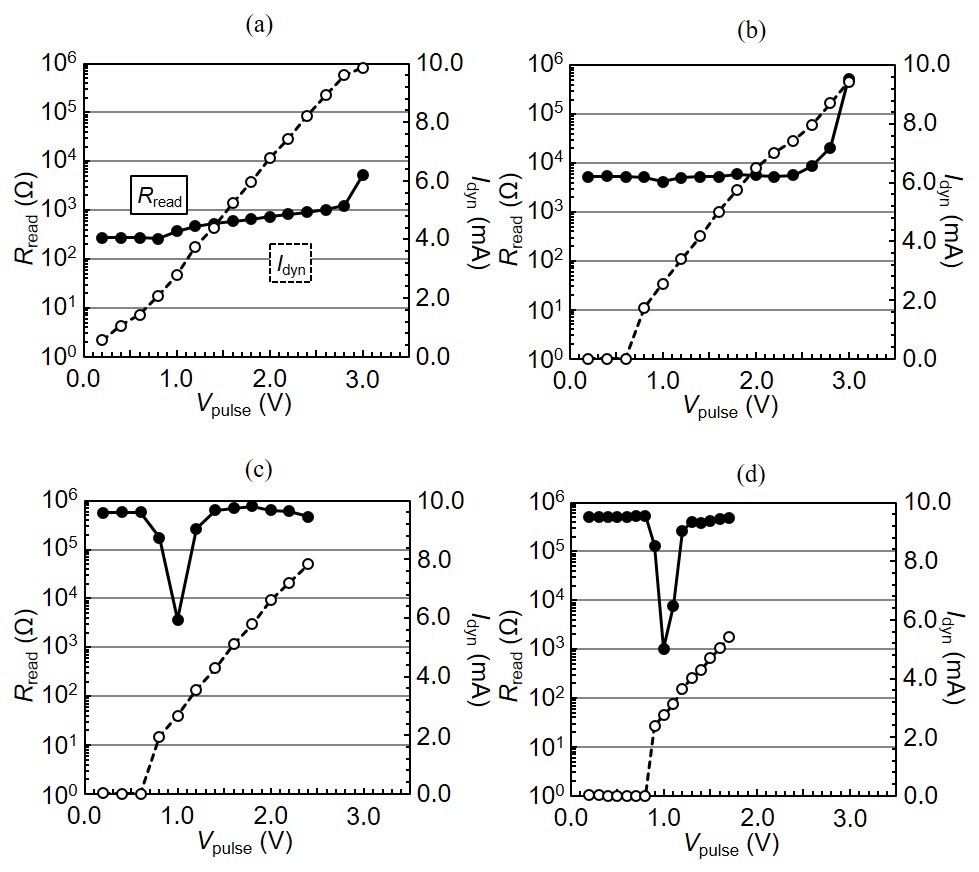}
  \caption{Electric properties in initialization of the Sn$_{0.5}$Te$_{0.5}$/Sb$_{2}$Te$_{3}$ SL material with low initial resistance through (a) 1st, (b) 2nd, (c) 3rd and (d) 4th pulse sweep.}
  \label{fig7}
 \end{center}
\end{figure}

As shown in Fig.\ \ref{fig4}, the initial resistances of the majority of the studied cells were high. However, some cells exhibited low initial resistances. Their electric properties are described in Fig.\ \ref{fig7}. According to the obtained results, the reset voltage {\it V}$_{rst}$ and {\it P}$_{rst}$ of the SnTeSL device are almost equal to those of GST, and then once the device is in the reset state, the device shows the low power switching. Thus, this process that converts the initial low resistance into the reset state can be called the initialization process for low power switching. Nevertheless, the described procedure has been rarely successful. In most cases, the cell was broken and exhibited extremely high resistance (such as 10$^{11}$ $\Omega$) during the reset process. The similar property was observed also in GeTeSL where high power is necessary to form the high resistance from the low initial resistance to start low power switching. Thus, this property seems common to SL materials. From these experimental facts, the high initial resistance seems essential in SL materials. This fact is not contradictory with the previous report\cite{simpson} where the initial resistance was high. Thus, this property seems to represent an important aspect of the switching mechanism of the tested SL devices, which will be discussed in more detail in Sec. \ref{discussions}.

\begin{figure}[hbt]
 \begin{center}
  \includegraphics[clip,width=2in]{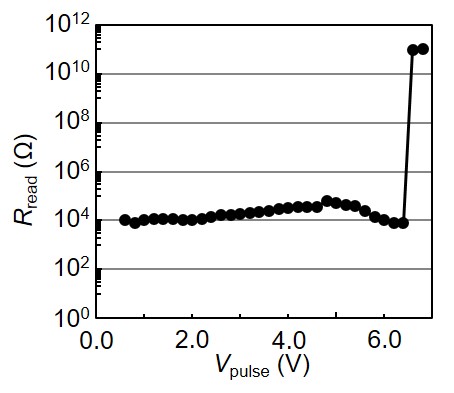}
  \caption{Relationship between the programmed voltage {\it V}$_{pulse}$ and the read resistance {\it R}$_{read}$ of the SnSbTe device, which consisted of the SnTe and Sb$_{2}$Te$_{3}$ layers.}
  \label{fig8}
 \end{center}
\end{figure}

Fig.\ \ref{fig8} displays the result of the pulse sweep measurement conducted for the SnSbTe device fabricated by sputtering Sn$_{0.5}$Te$_{0.5}$ and Sb$_{2}$Te$_{3}$ targets simultaneously at the sputtering rates equal to those utilized for the fabrication of the SnTeSL devices. They show that the tested device was broken before reaching the high resistance state. The likely reason for the absence of amorphization of the SnSbTe layers was their very high recrystallization rate due to the Sn-doped GST\cite{yamada2}\cite{song}. Hence, it can be concluded that the presence of a superlattice or layered structure significantly affects the resistance switching properties of SnTeSL.

\begin{figure}[hbt]
 \begin{center}
  \includegraphics[clip,width=3.4in]{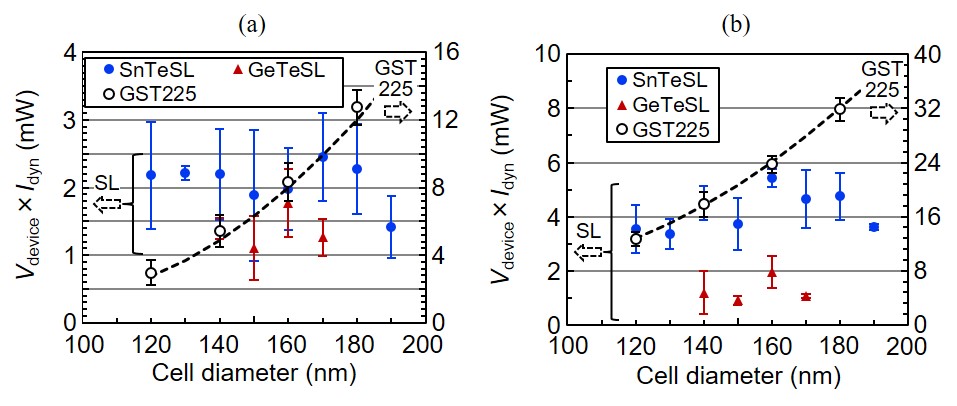}
  \caption{Scaling properties of the GST, GeTeSL, and SnTeSL materials determined during the (a) set and (b) reset operations.}
  \label{fig9}
 \end{center}
\end{figure}

Fig.\ \ref{fig9} displays the dependences of the set power {\it P}$_{set}$ and {\it P}$_{rst}$ of the GST, GeTeSL, and SnTeSL devices on their cell diameters (the scaling properties). About ten cells were examined for each device, and the obtained average values were plotted with the standard deviations drawn as the error bars. The obtained results show that both GeTeSL and SnTeSL devices exhibit no dependences on the cell diameter. Tai {\it et al.} has reported the scaling property of GeTeSL which concludes that the reset current shows approximately {\it r}$^{1.6}$ dependence where {\it r} represents the size of the resistor\cite{tai}. Here four points are discussed regarding the discrepancy between their conclusion and the data in this paper. One is the difference of the quantity of the vertical axis where one is the reset current and the other is the reset power. This difference, however, is not essential because the reset voltage {\it V}$_{rst}$ of SL devices showed no dependence of the cell size in this paper. Secondly, the number of the data points in this reference\cite{tai}  is too small to draw an exact fitting curve. Moreover, no error bars on which a fitting curve should pass are not shown. Thus, their conclusion might be somewhat hasty. The number of the data in this paper, however, is also not very large. In discussing this kind of properties statistically, the number of data should be more than tens or hundreds. This issue should be remained as a more precise future work. The third point to be discussed is the difference in configurations where the device in this reference\cite{tai}  is pillar while the devices in this paper have the configuration shown in Fig.\ \ref{fig1}. The effect brought about by the difference in configurations is, however, not clear because of the unclear mechanism of the resistance switching in SL devices. Lastly, the common point to this reference and this paper is, despite the quantitative difference, that the scaling properties of the SL devices are weaker than that of GST225, which seems relevant to the mechanism of resistance switching in SL devices. Though the exact conclusion must be drawn after more precise studies, the scalability will be discussed in Sec. \ref{discussions} under the assumption that the reset and set powers in the SL devices have almost no dependence on the cell size because Fig.\ \ref{fig9} seems correct for the SL devices (both GeTeSL and SnTeSL) with the configuration as shown in Fig.\ \ref{fig1}.

\begin{figure}[hbt]
 \begin{center}
  \includegraphics[clip,width=2in]{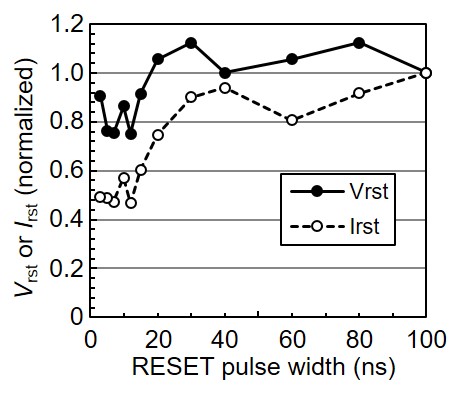}
  \caption{Relationships between the programmed reset pulse width and reset voltage ({\it V}$_{rst}$) / current ({\it I}$_{rst}$).}
  \label{fig10}
 \end{center}
\end{figure}

Fig.\ \ref{fig10} describes the reset properties of the studied cells determined at shorter pulse widths. Here, {\it V}$_{rst}$ is the value measured with the FET probe, which represents the voltage practically applied to the device. Because the short pulses become dull with time (see Fig.\ \ref{fig2}), their maximum amplitudes are plotted in Fig.\ \ref{fig10}. Both the voltage and current were normalized by the values of {\it V}$_{rst}$ and {\it I}$_{rst}$ measured at a pulse width of 100 ns, respectively. The {\it V}$_{rst}$ magnitudes obtained at different pulse widths were almost identical, while the values of {\it I}$_{rst}$ decreased after applying shorter pulses. As a result, the magnitude of {\it P}$_{rst}$ decreased with decreasing pulse width, which was also observed both qualitatively and quantitatively elsewhere\cite{shintani}, suggesting that this property was common for various SL structures.

\begin{figure}[hbt]
 \begin{center}
  \includegraphics[clip,width=2in]{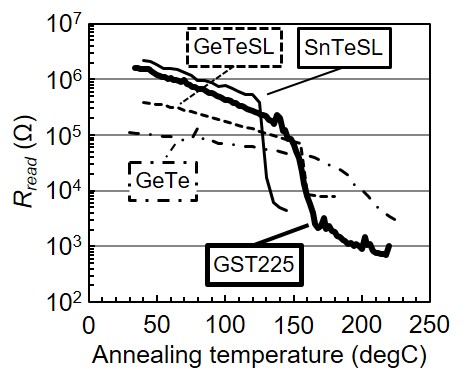}
  \caption{Relationships between the annealing temperature and resistances of the GST, GeTeSL, and SnTeSL materials, whose initial states corresponded to the reset states.}
  \label{fig11}
 \end{center}
\end{figure}

Fig.\ \ref{fig11} shows the dependence of the read resistance on the annealing temperature, indicating that the reset resistances of all devices rapidly decrease at temperature {\it T}$_{set}$. Furthermore, the {\it T}$_{set}$ of GeTeSL was almost equal to that of GST, although its value determined for the GeTe device was about 200$^{\circ}$C. From these results, it can be hypothesized that the interface between the GeTe and Sb$_{2}$Te$_{3}$ layers is mainly responsible for the resistance switching in the GeTeSL structure. This assumption can be used to explain the results presented in Fig.\ \ref{fig8} for SnTeSL.

\subsection{Properties of Sn$_{1-{\it x}}$Te$_{{\it x}}$/Sb$_{2}$Te$_{3}$ SL devices}
\label{Sn1-xTex}

\begin{figure}[hbt]
 \begin{center}
  \includegraphics[clip,width=3.4in]{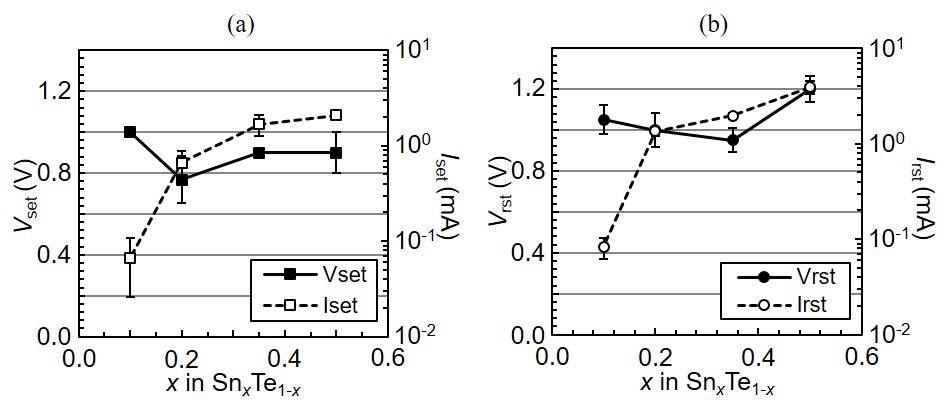}
  \caption{Set and reset voltages and currents measured for the Sn$_{x}$Te$_{1-x}$/Sb$_{2}$Te$_{3}$ materials at various {\it x} during the (a) set and (b) reset operations.}
  \label{fig12}
 \end{center}
\end{figure}

Fig.\ \ref{fig12} shows the set and reset voltages and currents of the Sn$_{x}$Te$_{1-x}$/Sb$_{2}$Te$_{3}$ SL devices, indicating that their {\it V}$_{rst}$ values obtained at various {\it x} were almost identical, while the magnitudes of {\it I}$_{rst}$ decreased at smaller {\it x}. Note that the right axis in this figure representing the current is logarithmic, while the left one denoting the voltage is linear. The measured {\it P}$_{rst}$ of the Sn$_{0.1}$Te$_{0.9}$/Sb$_{2}$Te$_{3}$ SL device was smaller than that of GST by a factor of less than 10$^{-3}$. Thus, the extreme reduction of the switching power at smaller {\it x} resulted mainly from the decrease in the switching current. This observation appears to be consistent with the results obtained after the application of short pulses in Fig.\ \ref{fig10}.

\begin{figure}[hbt]
 \begin{center}
  \includegraphics[clip,width=2in]{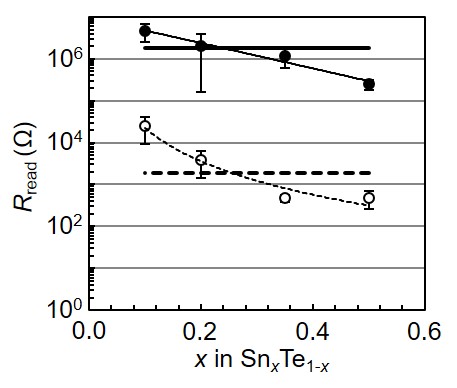}
  \caption{Set and reset resistances of Sn$_{x}$Te$_{1-x}$/Sb$_{2}$Te$_{3}$ as functions of {\it x}.}
  \label{fig13}
 \end{center}
\end{figure}

Fig.\ \ref{fig13} shows the set and reset read resistances of the Sn$_{x}$Te$_{1-x}$/Sb$_{2}$Te$_{3}$ SL devices, whose values increase with decreasing {\it x}. The {\it R}$_{rst,read}$({\it x})/{\it R}$_{set,read}$({\it x}) ratio between these two parameters decreases at smaller {\it x}; however, at {\it x} = 0.1, its value becomes equal to 10$^{2}$, which is sufficient for practical use.

\subsection{TEM images of SnTeSL devices}
\label{TEM}

\begin{figure}[hbt]
 \begin{center}
  \includegraphics[clip,width=3.4in]{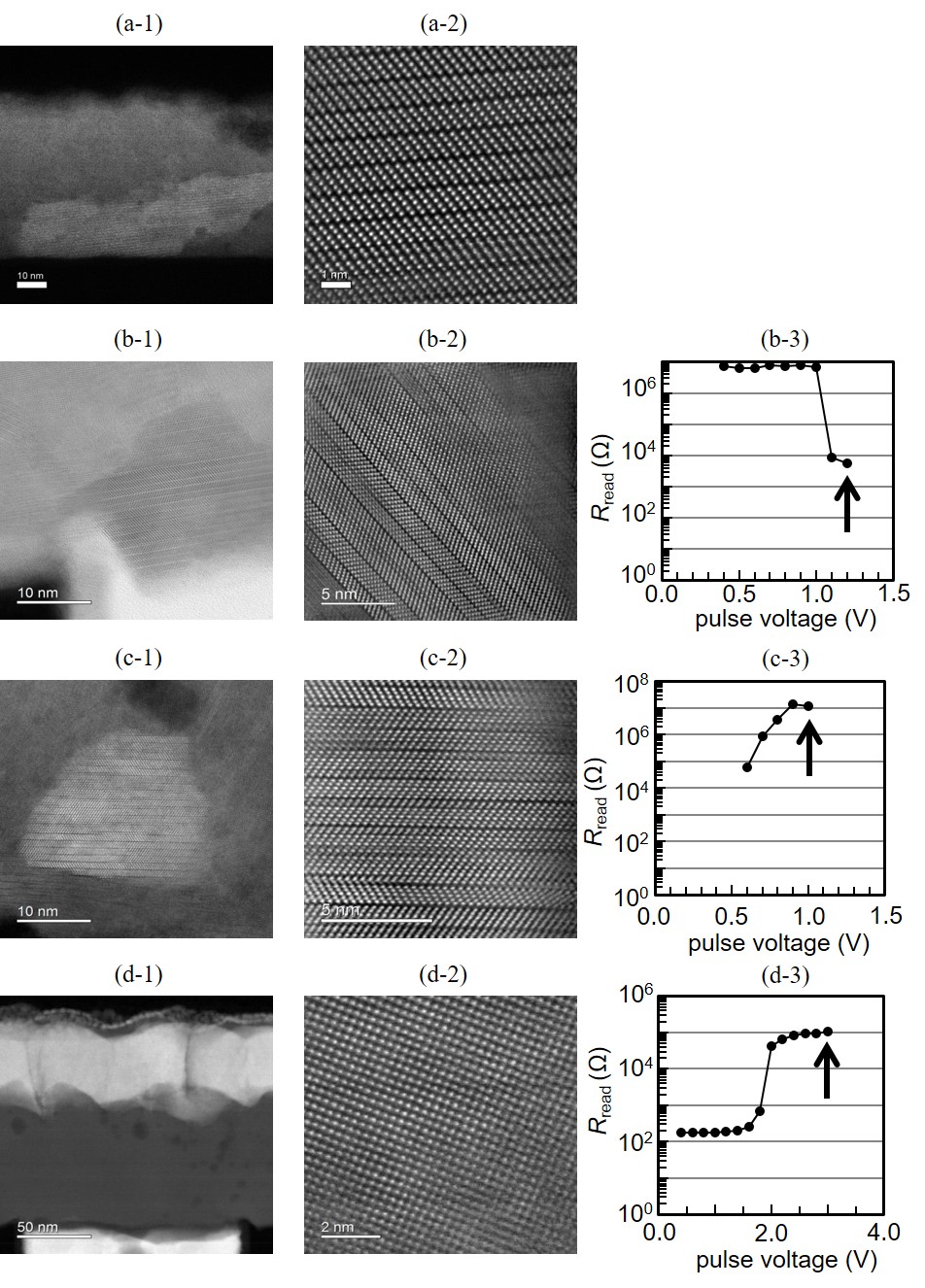}
  \caption{TEM images of Sn$_{0.5}$Te$_{0.5}$/Sb$_{2}$Te$_{3}$ obtained for the (a) initial state, (b) initialized set state, (c) reset state, and (d) cell broken by applying a high reset voltage. The left column (1) shows the macroscopic images, the middle column (2) contains the atomic scale images, and the graphs depicted in the right column (3) characterize the effects of the electric pulses applied to produce samples for TEM observations.}
  \label{fig14}
 \end{center}
\end{figure}

Fig.\ \ref{fig14} contains the TEM images of various states of the Sn$_{0.5}$Te$_{0.5}$/Sb$_{2}$Te$_{3}$ SL devices. In particular, Fig.\ \ref{fig14}(a), (b), and (c) show that the fabricated SL films are polycrystalline, although their grains exhibit similar structures characterized by the presence of five periodic atomic layers which likely originates from the Sb$_{2}$Te$_{3}$ layers with relatively long Te-Te bonding. The obtained three TEM micrographs exhibit no significant differences between the corresponding read resistances. This observation will be discussed in more detail in Sec. \ref{discussions}.

Fig.\ \ref{fig14}(d) shows the TEM images of the cells broken by applying a high voltage. Panel (d-3) demonstrates a higher {\it V}$_{rst}$ than that of panel (c-2), indicating that this cell was broken at least as a SL device although not completely because it exhibits resistance switching at voltage higher than the typical voltage. The sample depicted in Fig.\ \ref{fig14}(d-1) is characterized by a lower degree of crystallinity as compared to those of the specimens depicted in panels (a-1), (b-1), and (c-1). In contrast, the image presented in Fig.\ \ref{fig14}(d-2) does not exhibit the periodicity demonstrated in panels (a-2), (b-2), and (c-2). Thus, it can be concluded that good periodic properties are essential for achieving low {\it P}$_{rst}$ values.

Every TEM graph in Fig.\ \ref{fig14} shows the SL film in polycrystalline state though some regions show the periodic order. This might indicate that the well-aligned signle crystal superlattice structures are not always essential for low power switching. Fig.\ \ref{fig14}(a), (b) and (c) exhibit periodic structures consisting of five atoms while Fig.\ \ref{fig14} (d), which does not show low power switching, does not. These facts indicate, as circumstantial evidences, that this periodic structure must be the key for low power switching. Though Fig.\ \ref{fig14} shows the TEM graphs of Sn$_{0.5}$Te$_{0.5}$/Sb$_{2}$Te$_{3}$, the same discussion might be applied to Sn$_{0.1}$Te$_{0.9}$/Sb$_{2}$Te$_{3}$, which shows the lowest power switching in the superlattice phase change materials ever reported, because Soeya {\it et al.} reported the structure analyses of Sn$_{0.1}$Te$_{0.9}$/Sb$_{2}$Te$_{3}$ where the same periodic structures consisting of five atoms Te-(Sn, Sb)-Te-(Sn,Sb)-Te, are concluded\cite{soeya2}. Although this study challenged to identify the atomic species in this periodic layer, this identification failed because the atoms in this material have the close atomic numbers, 50, 51, and 52 where the identification using electron beams was impossible. Thus, the details of this structure are not clear. However, there is a possibility that Sn$_{x}$Te$_{1-x}$/Sb$_{2}$Te$_{3}$ studied in this paper have the same structure with the different composition ratio in (Sn, Sb). But still this hypothesis does not explain lower power switching for smaller {\it x} in Sn$_{x}$Te$_{1-x}$/Sb$_{2}$Te$_{3}$. The relationship between low power switching and the structures of superlattice phase change materials must be studied further.

\subsection{Superlattice cycles and switching power}
\label{cycle}

\begin{figure}[hbt]
 \begin{center}
  \includegraphics[clip,width=2in]{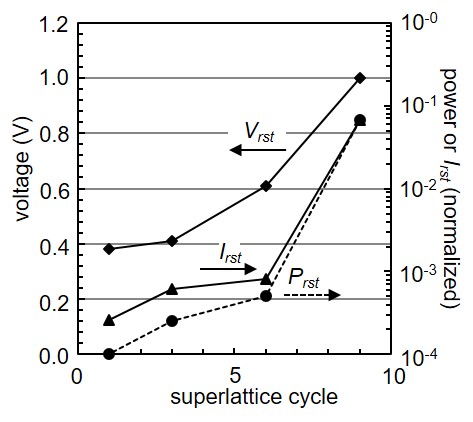}
  \caption{Dependence of {\it V}$_{rst}$, {\it I}$_{rst}$ and {\it P}$_{rst}$ on superlattice cycles in Sn$_{0.5}$Te$_{0.5}$/Sb$_{2}$Te$_{3}$ SL devices.}
  \label{fig15}
 \end{center}
\end{figure}

Fig.\ \ref{fig15} shows the dependence of {\it V}$_{rst}$, {\it I}$_{rst}$ and {\it P}$_{rst}$ on the number of the superlattice layers (superlattice cycles) in Sn$_{0.5}$Te$_{0.5}$/Sb$_{2}$Te$_{3}$ SL devices, where the nominal superlattice cycles described in Sec. \ref{Experimental} is 8. This figure shows that both {\it V}$_{rst}$ and {\it I}$_{rst}$ decrease as reduction of the superlattice cycles. This might be reasonable because the volume to be switched decreases as the superlattice cycles are reduced. Note, however, that the decrease of {\it I}$_{rst}$ is close to exponential while the decrease of {\it V}$_{rst}$ is close to linear. As a result, {\it P}$_{rst}$ becomes 10$^{-4}$ of that of GST in the device with the single superlattice cycle. The cause of this behavior will be discussed in Sec. \ref{discussions}.

\subsection{Multilevel cells of SL devices}
\label{multilevel}

MLCs are necessary components of the storage devices with large data capacities. Their formation can be achieved by applying a voltage of less than {\it V}$_{rst}$ to the PC device. For example, Fig.\ \ref{fig3}(a) shows that the voltages of 2.0 V, 2.2 V, and 2.6 V applied to GST resulted in the read resistances of 5.5${\times}$10$^{3}$ ${\Omega}$, 2.3${\times}$10$^{4}$ ${\Omega}$, and 1.4${\times}$10$^{5}$ ${\Omega}$, respectively. However, the formation of MLCs in SL devices appears to be a challenging task because their set and reset processes occur rapidly without a significant voltage margin, as shown in Fig.\ \ref{fig5}. The experiments performed using different cells show that applying a voltage of slightly less than {\it V}$_{rst}$ leads to a medium read resistance, which is too unstable for practical use.

\begin{figure}[hbt]
 \begin{center}
  \includegraphics[clip,width=3.4in]{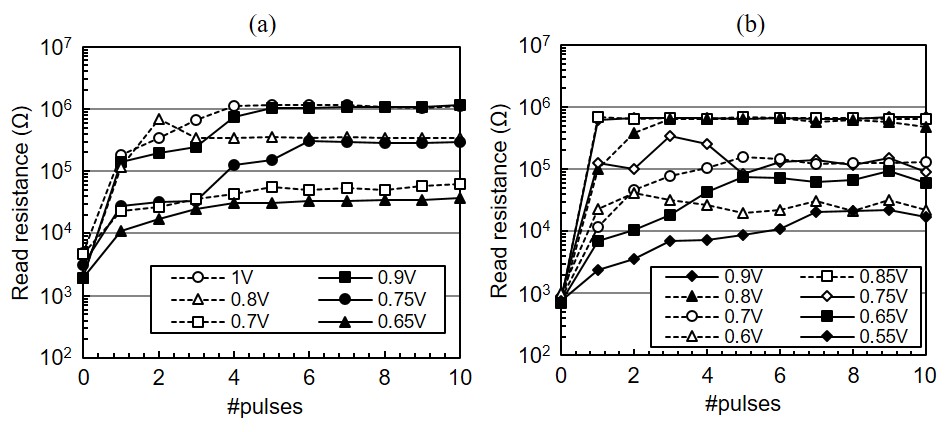}
  \caption{Relationships between the number of induced pulses and the read resistances of the (a) GeTeSL and (b) SnTeSL materials.}
  \label{fig16}
 \end{center}
\end{figure}

Fig.\ \ref{fig16} displays the results of multiple pulse testing aimed at mitigating this issue. It shows that both the GeTeSL and SnTeSL devices exhibit three-level discrete read resistances. One of the interesting characteristics in these figures is that some cases (“0.8 V” in Fig.\ \ref{fig16}(a) and “0.75 V” in Fig.\ \ref{fig16}(b)) show the resistances first increased and then decreased to constant values while other cases show the resistance increased with the number of pulses. This seems to indicate that these three levels are determined by some mechanism, which has not been fully clarified yet.

\section{Discussions}
\label{discussions}

This section mainly discusses the mechanism of low power switching in SL devices using partial switching and electric field-induced switching models. This task is performed by considering the data obtained in the current study. The explicability of each data by each model is listed in Table \ref{table}. The details on the models and their explicability are explained and discussed below.

\begin{table*}[hbt]
 \caption{Experimental results and interpretability of the proposed hypotheses. "Y" and "N" denote "explicability" and "inexplicability", respectively. (Y) indicates "conditional" or "uncertain".}
 \label{table}
 \centering
 \begin{tabular}{ccc}
  \hline
  Experimental data (figure) & Partial switching & Field-induced switching \\
  \hline \hline
  Initialization from high to low (Fig.\ \ref{fig4}) & Y & (Y) \\
  Initialization from low to high (Fig.\ \ref{fig7}) & Y & (Y) \\
  No switching in the alloyed state (Fig.\ \ref{fig8}) & Y & (Y) \\
  No scalability (Fig.\ \ref{fig9}) & Y & (Y) \\
  Low current switching by short pulses (Fig.\ \ref{fig10}) & N & Y \\
  Annealing properties (Fig.\ \ref{fig11}) & Y & (Y) \\
  Low current switching in Sn$_{x}$Te$_{1-x}$/Sb$_{2}$Te$_{3}$ (Fig.\ \ref{fig12}) & Y & Y \\
  Low power switching in layered structure (Fig.\ \ref{fig14}) & Y & Y \\
  Reduction in {\it P}$_{rst}$ with decreasing SL cycles (Fig.\ \ref{fig15}) & N & Y \\
  MLC formation by multiple pulses (Fig.\ \ref{fig16}) & N & N \\
  \hline
 \end{tabular}
\end{table*}

\begin{figure}[hbt]
 \begin{center}
  \includegraphics[clip,width=3.4in]{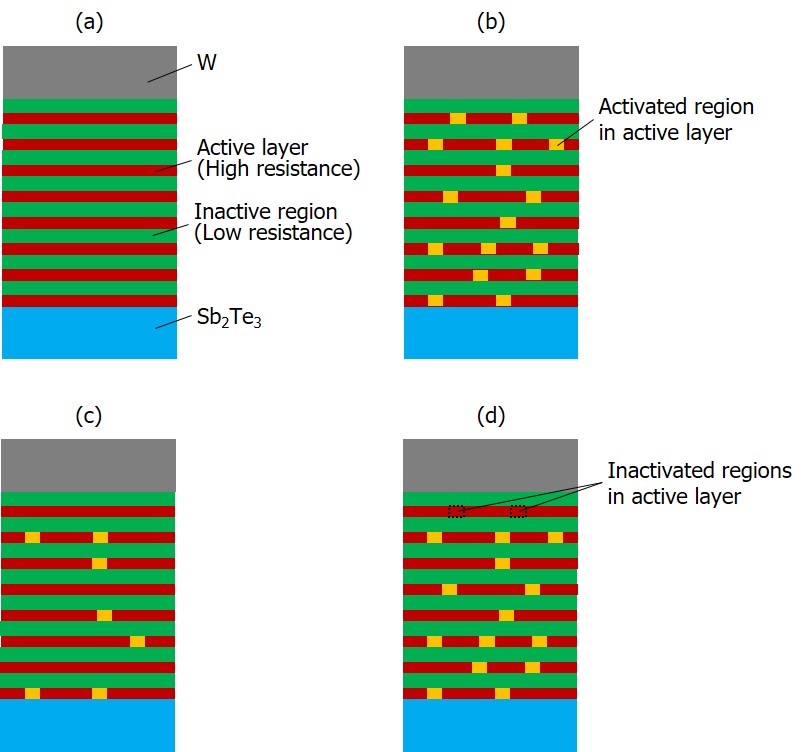}
  \caption{Diagrams illustrating the partial switching model: (a) initial (as-deposited), (b) initialized, (c) reset, and (d) unsuccessfully set states.}
  \label{fig17}
 \end{center}
\end{figure}

The diagrams describing the partial switching model are shown in Fig.\ \ref{fig17}. Here the activated regions are layered, which is essential for low power switching as shown in the TEM graphs depicted in Fig.\ \ref{fig14}. Fig.\ \ref{fig17}(a) displays the initial as-deposited state characterized by relatively high resistance. The initialization process produces multiple low resistance regions (Fig.\ \ref{fig17}(b)), which represent activated regions for the overwriting operation, while the other regions retain their high resistances. The reset operation switches the resistance of some of these activated regions (Fig.\ \ref{fig17}(c)) because it requires switching all the activated regions of only one layer into the high resistance state.

This partial switching model can satisfactorily explain the data presented in Fig.\ \ref{fig4}, Fig.\ \ref{fig7}, Fig.\ \ref{fig8}, Fig.\ \ref{fig9}, Fig.\ \ref{fig11}, and Fig.\ \ref{fig12}. The initialization processes in Fig.\ \ref{fig4} and Fig.\ \ref{fig7} can occur partially in some regions since the inhomogeneity in the deposited films produces sites that are relatively easy to switch (easily-switching sites). When these sites are switched multiple times during overwriting, the electric properties of the cell remain stable. The results presented in Fig.\ \ref{fig7} can be explained as follows. If the initial resistance of the cell is low, the switching process requires relatively high power (similar to that used in ordinary phase change memory chips). Once the resistance of the entire cell becomes high, low power switching sites must be generated in the next step. The result in Fig.\ \ref{fig8} (no switching in the alloyed state) is also explained if it is assumed that the layered structure or the interfaces of the layers generate the easily-switching sites. Furthermore, the absence of scalability in Fig.\ \ref{fig9} is also explained because the switching power of a material is typically determined not by the cell size, but by the number of switching sites. The annealing properties of the studied materials described in Fig.\ \ref{fig11} can be elucidated by assuming that the switching mechanism corresponds to a conventional phase change reaction consisting of the crystallization and amorphization processes at the switching sites. The low current switching depicted in Fig.\ \ref{fig12} can be explained by modeling the switching sites as parallel resistors. In this case, the electric current is reduced by decreasing the number of the resistors, while the voltage remains almost the same. The low power switching in the layered structure (TEM graphs in Fig.\ \ref{fig14}) can be explained by assuming that the switching sites are located at the layer interfaces. For example, in the GeTeSL structure, both the GeTe and Sb$_{2}$Te$_{3}$ layers do not exhibit low power switching properties, while their alloying ratio varies in the vertical direction. If some interfacial sites are characterized by the optimal alloying ratio, they can undergo low power switching.

This partial switching model might explain the cause of the endurance properties in Fig.\ \ref{fig6} by, for example, the following hypothesis as shown in Fig.\ \ref{fig17}(d). If the atomic mobility in some of the activated regions becomes low because of a certain local condition, the difference between the set and reset resistance becomes low until new regions are activated. This phenomenon is similar to resistive random access memories (ReRAMs) where similar instability in endurance is often observed\cite{huang}\cite{arita}. If the switching mechanism of SL materials is similar to that of ReRAMs, the instability of endurance in SL devices might be solved by the similar circumventions adopted in ReRAMs such as surface modification\cite{huang} and so on.
Although the partial switching model can explain many experimental findings as was shown above, it is not consistent with the data presented in Fig.\ \ref{fig10}, Fig.\ \ref{fig15} and Fig.\ \ref{fig16}. While it was possible to elucidate the results of Fig.\ \ref{fig16} in terms of switching a limited number of the switching sites induced by the low voltage, the generated read resistances exhibit exponential behavior, suggesting that the number of the switched sites was exponentially dependent on the applied voltage, which appeared to be explicable with the conventional theories of phase change. Moreover, this model requires that the total volume of the switching sites in the Sn$_{0.1}$Te$_{0.9}$/Sb$_{2}$Te$_{3}$ SL structure be equal to about 10$^{-3}$ of that in the Sn$_{0.5}$Te$_{0.5}$/Sb$_{2}$Te$_{3}$ SL structure, which is relatively hard to maintain from one experiment to another. However, the obtained data exhibit high repeatability, as indicated by the error bars in Fig.\ \ref{fig12}. Thus, the partial switching model is unable to explain all the phenomena observed in this work.

Now the other model, the electric field-induced switching model which is capable of explaining almost all the data obtained in this study (except for Fig.\ \ref{fig16}), is considered. The results depicted in Fig.\ \ref{fig4} and Fig.\ \ref{fig7} are interpreted using an approach similar to that of the partial switching model if it is assumed that the electric field can switch the resistance in SL cells with some mechanism. No switching in the alloyed state (Fig.\ \ref{fig8}) can be explained if the utilized model is applicable only to the layered structure. 

\begin{figure}[h]
 \begin{center}
  \includegraphics[clip,width=3in]{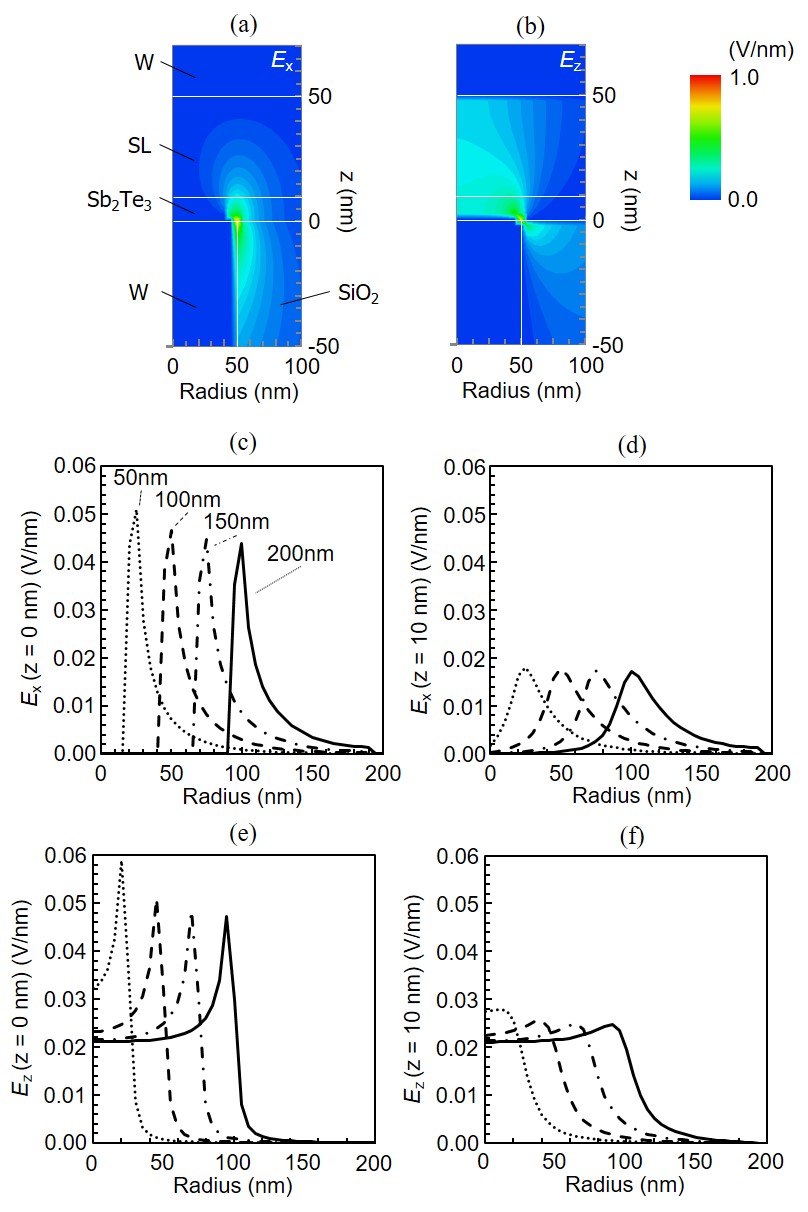}
  \caption{Electric field distributions determined computationally by applying a voltage of 1 V between the bottom and top electrodes of the cell. The {\it x} and {\it z} axes are oriented along the horizontal and vertical directions, respectively. Electric field components (a) {\it E}$_{x}$ and (b) {\it E}$_{z}$ of the cell with a diameter of 100 nm. Electric field components (c) {\it E}$_{x}$ ({\it r}, {\it z} = 0 nm), (d) {\it E}$_{x}$ ({\it r}, {\it z} = 10 nm), (e) {\it E}$_{z}$ ({\it r}, {\it z}= 0 nm), and (f) {\it E}$_{z}$ ({\it r}, {\it z} = 10 nm) of the cells with diameters of 50, 100, 150, and 200 nm.}
  \label{fig18}
 \end{center}
\end{figure}

The data presented in Fig.\ \ref{fig10}, Fig.\ \ref{fig12} and Fig.\ \ref{fig15}, which show that low power switching is caused by the extremely low electric current during the pulse incidence, can be satisfactorily explained by this model utilizing only the electric voltage or field as the main physical parameter and not the electric current nor the power such as Joule's heat. In this case, the annealing properties described in Fig.\ \ref{fig11} can be elucidated by assuming that the resistance switching induced by the applied electric field represents a conventional phase change process consisting of the crystallization and amorphization stages. Furthermore, the absence of scalability in Fig.\ \ref{fig9} may be explained by estimating the intensity of the electric field induced by applying voltage to the cell. Its calculated values are shown in Fig.\ \ref{fig18}. The computation was performed in the cylindrical coordinates with the {\it x} and {\it z} axes defined as the horizontal and vertical directions, respectively. In these calculations, the resistivities of the Sb$_{2}$Te$_{3}$ and GeTe layers were assumed to be equal to that of crystalline GST, which did not significantly affect the obtained results because the resistivities of crystalline GST, Sb$_{2}$Te$_{3}$, and GeTe were about three orders of magnitude higher than that of W. The TiN underlayer was ignored because of its very small thickness. 

The results presented in Fig.\ \ref{fig18} show that both {\it E}$_{x}$ and {\it E}$_{z}$ are localized around the edge of the electrode and {\it E}$_{x}$ is more localized than {\it E}$_{z}$. The experimental data reported elsewhere\cite{shintani} indicate that the parameter {\it E}$_{x}$ mainly affects the low power switching process. From the results presented above, it can be concluded that the switching power is independent of the cell size because the driving force {\it E}$_{x}$ is localized at the cell edge. Moreover, Fig.\ \ref{fig18}(c)–(f) show that the peak intensity of the field depends only weakly on the cell size, especially at {\it z} = 10 nm where the SL structure begins. These results are consistent with the TEM graphs presented in Fig.\ \ref{fig14}, indicating that the vectorial parameters (such as electric field) produce a greater effect on the layered structure than on the bulk system. The similar phenomenon has been reported\cite{yang} where the atoms in GST migrate along the external electric field (Ge and Sb to the cathode and Te to the anode). This phenomenon might be related to the mechanism of resistance switching in SL materials. For example, Sb and Te migrate to the opposite sides, which should affect the behavior of the van der Waals forces at Te-Te bonds in Sb$_{2}$Te$_{3}$ layers. If the external electric field reduces the distance of Te-Te bonds and applies the opposite forces to Ge and Te atoms, GeTe might get easier to change its phase, which leads to low power switching. This phenomenon easily occurs in SL materials maybe because the atoms are aligned more periodically in SL materials than in GST even though SL materials are in polycrystalline states. This discussion is applied also to SnTeSL because Sn is in the same group as Ge in the periodic table and has the similar ionic properties. A non-thermal phase change has been observed previously by applying picosecond pulses\cite{huang2}. Hence, it can be concluded that low power switching was achieved in this work due to the application of nanosecond pulses because the total amount of {\it E}$_{x}$ was greater in the layered structure despite the polycrystallinity of the SL layers (see Fig.\ \ref{fig14}). These data can qualitatively explain the non-scalability properties depicted in Fig.\ \ref{fig9}. The {\it E}$_{x}$ component is localized along the perimeter of the cell, which lead to the conclusion that the scalability of the studied cells is characterized not by their radii, but by the linear dependence of the radii. However, this conclusion is not consistent with the results presented in Fig.\ \ref{fig9}, which may be explained by the combination of the partial switching and electric field-induced switching models, where the applied electric field switches the resistance only at certain regions of the cell.

Although the electric field-induced switching model can partially explain the obtained experimental data, the results presented in Fig.\ \ref{fig16} are not fully consistent with it. Moreover, many particular details of the related mechanism are not clear; thus, the explanation of some results obtained using this model is only "conditional".

While neither of these two models can explain all the experimental data obtained in this work consistently, some aspects did become clearer. One of them is that the existence of a layered or superlattice structure is essential for low power switching. However, a particular role of this structure has not been fully clarified yet. Various factors can be considered here such as the interfaces between multiple layers or the internal stress generated by the stratification of different materials.

The roles of the internal stress in phase change phenomenon have been reported in some previous reports\cite{kalikka}\cite{zhou}\cite{roscioni}\cite{eising}. Though these studies focus on crystallization from the amorphous phase whose mechanism might not be applied to SL materials, similar discussions might be applied in common from the view point of the atomic mobility. It has been reported that the switching power decreases with decreasing number of SL layers, and its extremely low value (about 10$^{-4}$ of that of GST225) is observed for single-layered SnTeSL as shown in Fig.\ \ref{fig15}. It can be thought that the internal stress is higher in a thinner heterostructure film because a thinner film cannot relax the internal stress sufficiently. However, since no direct stress measurements were obtained to support any of these hypotheses, additional studies in this area must be performed.

The effect of the interfaces has also been discussed. Simpson {\it et al.}\cite{simpson} suggested that the interface between the GeTe and Sb$_{2}$Te$_{3}$ layers produced the greatest effect, whereas Ohyanagi {\it et al.}\cite{ohyanagi3} stressed the importance of the top layer of the SL material. Since no results were obtained to support any of these hypotheses, additional studies in this area must be performed.

As shown in Table \ref{table}, no consistent model fully describing the resistance switching process in phase change SL materials has been developed yet. Hence, either a combination of the two models discussed above or a completely different mechanism must be considered. Additional macroscopic and microscopic studies are required for elucidating the resistance switching mechanism in SL materials and their practical use for manufacturing low power switching devices.

Next, the difference between GeTeSL and SnTeSL is discussed. Though the mechanism of the resistance switching in SL devices is still unclear, it might be thought that the same principle works for both SL materials because the experimental results shown in Fig.\ \ref{fig9}, Fig.\ \ref{fig10} and Fig.\ \ref{fig16}, which exhibit the essential differences from GST225, are qualitatively common to both SL materials. Recently the effect of van der Waals gaps in GST or GeTeSL has been discussed in relation to low power switching in GeTeSL\cite{lotnyk}\cite{kolobov2}\cite{chen}. These discussions might be applicable also to SnTeSL because, as already shown and discussed, the TEM images in Fig.\ \ref{fig14} show that the devices which show low power switching contain the vacancy layers which are considered as van der Waals gaps while this vacancy layer is not observed in the broken device which does not exhibit low power switching. Thus, as in GeTeSL devices, this vacancy layer (possibly van der Waals gap) might possibly be the essential cause of low power switching. If van der Waals gap is modulated easily by the electric field, the electric field induced model can be supported. The vacancy layer observed in SnTeSL, however, has not been proved to be van der Waals gap. Moreover, it is controversial whether the multilevel resistances shown in Fig.\ \ref{fig16} can be explained in terms of van der Waals gap. Further studies are necessary from these points of view.

Lastly the perspectives of phase change SL materials are discussed. Though this paper has reported some advantages of SnTeSL material, there remain some issues to be solved, especially i) the instability of the unstable endurance property (Fig.\ \ref{fig6}), ii) the variation of the initial resistance (Fig.\ \ref{fig7}), iii) the retention property predicted by the thermal property (Fig.\ \ref{fig11}), and iv) the power consumed for resistance switching in smaller cells in connection with non-scalability (Fig.\ \ref{fig9}). Though the issues i) and ii) will possibly be solved by the operations such as an initialization algorithm and the adjustment of the formed resistance by verification, it is more desirable for practical use to solve these issues because these operations would take time and electric power to store data. Though the causes of these properties are not clear, they are thought to be related to the conditions of the as-deposited films in the device, which have to be solved by the device design and/or the fabrication processes. One of the possibilities of the cause is the polycrystalline state as shown in Fig.\ \ref{fig14}. If the initial state was in the well-aligned single crystalline state, there would be no defects which would lead to the initial high resistance and the good endurance property due to the well-controlled generation of the activated regions. Moreover, as already discussed above, these issues might be solved by, for example, controlling the boundary between the electrodes and the SL material as circumvented in ReRAMs\cite{huang}. The issue iii) might be solved by, for example, adding some impurity which increases the thermal stability. Addition of impurity, however, increases the reset power in many cases. The atomic species of the impurity which can fulfill these two requirements should be selected carefully, where a detailed clarification of the mechanism of resistance switching in the SL devices will be of great importance. Regarding the issue iv) on the scalability, the property of non-scalability seems to decline the value of SL materials. SnTeSL, however, showed the switching power of 10$^{-3}$ – 10$^{-4}$ compared with GST225 using the cell with the diameter of 100 nm. This means that, by assuming that the switching power is simply proportional to the cell area, the switching power of SnTeSL will be 10$^{-1}$ – 10$^{-2}$ of that of GST225 if the cell diameter is 10 nm with the higher endurance than GST225. This is still highly advantageous for semiconductor memory industry. However, as described above, one of the serious issues is the retention property. SnTeSL should be investigated further to solve this issue for practical use in the future.

\section{Conclusions}
\label{conclusions}

The fundamental properties of the phase change Superlattice (SL) materials, mainly Sn$_{x}$Te$_{1-x}$/Sb$_{2}$Te$_{3}$ SL, were investigated in this work. It was found that their low power switching required a high initial resistance. The switching power of Sn$_{x}$Te$_{1-x}$/Sb$_{2}$Te$_{3}$ SL decreased with decreasing {\it x} and was equal to about 10$^{-3}$ of the value obtained for the Ge$_{2}$Sb$_{2}$Te$_{5}$ (GST225) conventional phase change material at {\it x} = 0.1. During the pulse incidence, the electric current was drastically reduced, while the switching voltage remained almost equal. The results of experiments conducted using short pulses revealed that a lower electric current was required for switching, while the switching voltage did not depend on the pulse width. The TEM images and experimental data obtained using the alloyed SnTe and Sb$_{2}$Te$_{3}$ layers indicate that the presence of a superlattice or a layered structure is essential for low power switching. At the same time, no dependence of the switching power on the cell size (scaling) was observed. Although the electric properties of the SL materials are different from those of the conventional GST materials, their annealing properties are very similar. In addition, multilevel cells were obtained by performing a multi-pulse reset operation. The partial switching and electric field-induced switching (as already proposed elsewhere\cite{shintani}) models were proposed to explain the experimental results obtained in this study, although none of them was able to explain the experimental results perfectly. Hence, additional studies are required to clarify the mechanism of the resistance switching in SL materials.

\section*{Acknowledgements}
\label{acknowledgements}

The author expresses his gratitude to Dr.\ Junji Tominaga from the Advanced Institute of Science and Technology (AIST) for providing necessary materials and fruitful discussions. He would also like to acknowledge Dr.\ Susumu Soeya and Ms.\ Reiko Kondo from AIST for sample preparation. He would like to express the deepest appreciation to Dr.\ Noboru Yamada for his constant and considerable encouragement. This research was supported by the Japan Society for the Promotion of Science (JSPS) through the "Funding Program for World-Leading Innovative R\&D on Science and Technology (FIRST)" initiated by the Council for Science and Technology Policy (CSTP).

\end{document}